\documentstyle[aps,prd,preprint]{revtex}

\begin{document}

\draft
\title{Probing negative dimensional integration:\\
two-loop covariant vertex and one-loop light-cone integrals}
\author{A. T. Suzuki, A. G. M. Schmidt, and R. Bent\'{\i}n}
\address{Universidade Estadual Paulista -- Instituto de F\'{\i}sica Te\'orica, R.Pamplona, 145, S\~ao Paulo SP
CEP 01405-900, Brazil }
\date{\today}

\maketitle

\begin{abstract} 
Negative dimensional integration method (NDIM) seems to be a very promising
technique for evaluating massless and/or massive Feynman diagrams. It is unique
in the sense that the method gives solutions in different regions of external
momenta simultaneously. Moreover, it is a technique whereby the difficulties
associated with performing parametric integrals in the standard approach are
transferred to a simpler solving of a system of linear algebraic equations,
thanks to the polynomial character of the relevant integrands. We employ this
method to evaluate a scalar integral for a massless two-loop three-point vertex
with all the external legs off-shell, and consider several special cases for
it, yielding results, even for distinct simpler diagrams. We also consider the
possibility of NDIM in non-covariant gauges such as the light-cone gauge and do
some illustrative calculations, showing that for one-degree violation of
covariance (i.e., one external, gauge-breaking, light-like vector $n_{\mu}$)
the ensuing results are concordant with the ones obtained via either the usual
dimensional regularization technique, or the use of principal value
prescription for the gauge dependent pole, while for two-degree violation of
covariance --- i.e., two external, light-like vectors $n_{\mu}$, the
gauge-breaking one, and (its dual) $n^{\ast}_{\mu}$ --- the ensuing results are
concordant with the ones obtained via causal constraints or the use of the
so-called generalized Mandelstam-Leibbrandt prescription.

\end{abstract}

\def\be{\begin{equation}}
\def\ee{\end{equation}}
\def\beq{\begin{eqnarray}}
\def\eeq{\end{eqnarray}}
\def\s{\sigma}
\def\G{\Gamma}
\def\F{_2F_1}
\def\an{analytic}
\def\ac{\an{} continuation}
\def\hsr{hypergeometric series representations}
\def\hf{hypergeometric function}
\def\ndim{NDIM}
\def\quarto{\frac{1}{4}}

 \def\half{\frac{1}{2}}

\section{Introduction.}

The equivalence between fermionic integration, in positive dimensional
space-time, and bosonic integration, in negative dimensional space-time, is
a remarkable property\cite{halliday2}. Based on this fact, Halliday and
co-workers suggested the use of such a property to tackle the problem of
calculating Feynman integrals. The advantage to be derived in such an
approach can be appreciated and understood --- at least in principle --- from
the point of view that Grassmannian integrals are linear operators defined with
few rules and properties and thus easier to solve them than the ordinary
integrals. 

Negative dimensional integration method (NDIM) allows us to perform massless
Feynman integrals with propagators raised to arbitrary powers even at the
two-loop level\cite{easy}. Box diagrams with massive propagators can also be calculated
easily and several results are obtained according to the different
possibilities of dimensionless ratios defined by internal mass and external
momenta\cite{box}. This work is intended to advance our testing to the next
natural step at the two-loop level  of three-point vertex diagrams in the
framework of off-shell external momenta. Furthermore, recalling that the
light-cone gauge loop integrals are notoriously more difficult to handle than
their covariant counterparts in virtue of the structure of the gauge boson
propagator\cite{leib}, we show how \ndim{} can simplify things out even in this
case. 

So, our proposal here is to give clear-cut examples to demonstrate the beauty
and power of this technique in dealing with Feynman integrals of the
perturbative quantum field theories. Without loss of generality, here we
restrict ourselves to massless fields. The outline for this work is given as
follows: In Section II we solve a scalar integral for the diagram of Fig.1 and discuss some special cases stemming from it. In Section III we consider a
scalar and a vectorial one-loop light-cone integrals pertaining to two-point
self-energy diagrams and finally in the last Section we present our concluding
remarks.

\section{Off-shell two-loop three-point vertex.}

This computation is performed following a few simple steps outlined in
\cite{lab,flying}. First of all, let us consider the Gaussian-like integral,

\beq 
\label{gauss} 
I &=& \int\int d^D\!r\; d^D\!q \;\;\exp\left[-\alpha q^2 -\beta
(q-p)^2 -\gamma r^2 -\xi (q-r-k)^2\right]\nonumber\\
&=&  \left(\frac{\pi^2}{\phi}\right)^{D/2} \exp{\left[ -\frac{1}{\phi}\left(
\alpha\gamma\xi k^2+ (\alpha\beta\gamma +\alpha\beta\xi)p^2+
\beta\gamma\xi t^2\right)\right]} ,
\eeq
where the arguments in the exponential function of the integrand correspond to
propagators in the diagram of Fig. 1 and for compactness we define
$\phi=\alpha\gamma+\alpha\xi+\beta\gamma+\beta\xi+\gamma\xi$. The arbitrary
parameters $(\alpha,\beta,\gamma,\xi)$ are chosen such that their real parts
are positive to make sure we have well-defined objects over the whole
integration space.

Expanding $I$ in Taylor series we obtain,

\be 
I = \sum_{i,j,l,m=0}^\infty \frac{(-1)^{i+j+l+m}}{i!j!l!m!}\alpha^i
\beta^j\gamma^l\xi^m {\cal S}_{NDIM},
\ee
where ${\cal S}_{NDIM}$ is the relevant integral in negative $D$, defined by 

\be 
\label{Indim} {\cal S}_{NDIM} = \int\int d^D\!q\;d^D\!r
\;(q^2)^i\left[(q-p)^2\right]^j (r^2)^l\left[(r-q+k)^2\right]^m . 
\ee

Now comparing both expressions for the original integral $I$ we are led to the
conclusion that in order to have equality between these expressions, the factor
${\cal S}_{NDIM}$ must be given by the multiple series,

\be 
\label{geral} 
{\cal S}_{NDIM} = {\cal G}(i,j,l,m;D) \sum_{n_1,...,n_9=0}^\infty
\frac{(p^2)^{n_1+n_2} (k^2)^{n_3}(t^2)^{n_4}} {\prod_{i=1}^9 n_i!}
 \; \delta_{a,i}\;
\delta_{b,j}\;\delta_{c,l}
\;\delta_{d,m} ,
\ee
where $(a,b,c,d)$ stand respectively for

\beq
a&=&n_1+n_2+n_3+n_5+n_6\:,\nonumber\\
b&=&n_1+n_2+n_4+n_7+n_8\:,\nonumber\\
c&=&n_1+n_3+n_4+n_5+n_7+n_9\:,\nonumber\\
d&=&n_2+n_3+n_4+n_6+n_8+n_9\:,
\eeq
with 
\[ {\cal G}(i,j,l,m;D) = (-\pi)^D\G(1+i)\G(1+j)\G(1+l)\G(1+m)\G(1-\s-\half D),
\] 
and for convenience we use the definition $\s=i+j+l+m+D$. The system we must
solve is, therefore,

\be 
\label{sys2}\left \{ \begin{array}{rcl}
n_1+n_2+n_3+n_5+n_6&=&i \\[.25cm]
n_1+n_2+n_4+n_7+n_8&=&j \\[.25cm]
n_1+n_3+n_4+n_5+n_7+n_9&=&l \\[.25cm]
n_2+n_3+n_4+n_6+n_8+n_9&=&m \\[.25cm]
n_1+n_2+n_3+n_4&=&\s . \end{array} \right. 
\ee

It is an easy matter to see that this system is composed of five equations with
nine ``unknowns'' (the sum indices), so that it cannot be solved unless it is
done in terms of four arbitrary ``unknowns''. These, of course, will label the
four remnant summations, which means that the answer will be in terms of a
fourfold summation series. From the combinatorics, it is a straightforward
matter to see that there are many different ways we can choose those four
indices; indeed, we can choose $C_5^9=126$ different ways. In other words, what
we need to do is to solve $126$ different systems. Of these, $45$ are
unsolvable systems, i.e., they are systems whose set solution is empty. There
remains therefore $81$ which have non-trivial solutions. Of course, the trivial
solutions are of no interest at all. However, the non-trivial solutions
generate a space of functions with different basis, characterized by their
functional variable, according to the different possibilities allowed for
ratios of external momenta. Each basis is a solution for the pertinent Feynman
integral, which is connected by \ac{} to all other basis defined by other set
solutions. We remind ourselves that a basis that generates a given space can be
composed of one or several linearly independent functions combined in what is
called linear combination.

Each representation of the Feynman integral will be given by a basis of
functions generated by the solutions of the systems\cite{suzuki2}. Of course,
only linearly independent and non-degenerate solutions are relevant to define a
basis.

It can be easily seen that the diagram we are dealing with here is symmetric
under the exchange of external momenta $k^2 \leftrightarrow t^2$. This symmetry
is reflected by the systems we have to solve, and the solutions display this
fact. Moreover, in order to save ourselves space, only those solutions which
are not thus connected have been written down explicitly (see \cite{preprint}).
Here we further restrict ourselves to those solutions which are of present
interest, that is, the ones we can compare with results obtained from positive dimensional techniques.

With these clarifying statements and with the solutions of the many systems in
hands --- it is an easy matter to write down a computer program to solve all
the systems --- and the general form of the results (\ref{geral}), we can start
building the power series representations for the Feynman graph.

We begin our analysis of the solutions for the systems by looking at the
simpler ones, i.e., those having two variables, defined by ratios of external
momenta. Of course, there are in fact fours sums but two of them have unity
argument, making it possible for us to actually sum the pertinent series as we
shall shortly see. The variables are $(x,y) ,\;\; (z,y^{-1}),\;\;
(x^{-1},z^{-1})$ where we define the dimensionless ratios

\be
x=\frac{p^2}{k^2},\qquad\quad
y=\frac{t^2}{k^2},\qquad\quad
z=\frac{p^2}{t^2}.
\ee

Each of the three pairs above appears twelve times among the total of $81$
systems with non-trivial solutions. Yet, each of these is twelve-fold
degenerate just as in the on-shell case calculated in \cite{lab}. A possible way
of expressing the first solution in positive $D$ --- the \ac{} to this region
is carefully explained in \cite{lab,flying} --- is:

\beq 
\label{S1} S_1^{AC} &=& \pi^D(p^2)^\s P_1^{AC}\nonumber\\[.25cm]
& \times & F_4\left(-\s, -l-m-\half D; 1+j-\s,
1+i-\s\left|x^{-1},z^{-1}\right.\right),
\eeq 
with $P_1^{AC}$ being given by
\beq 
P_1^{AC} &=& (-i|\s)(-j|\s)(-l|-m-\half D)(-m|l+m+\half D) \nonumber\\[.25cm]
& \times &(l+m+D|-l-\half D)(\s+\half D|-2\s-\half D) ,
\eeq
and where we use the Pochhammer symbol $(a|k) \equiv (a)_k = \G(a+k)/\G(a)\;$,
together with one of its properties, $(a|b+c) = (a+b|c)(a|b)$ as well as the
well-known Gauss' summation formula \cite{bateman} for the \hf{} $\F$
with unit argument, 
\be 
\label{2f1} \F(a,b;c|1) = \frac{\G(c)\G(c-a-b)}{\G(c-a)\G(c-b)}
,
\ee 
so that two of the four series above are summed up. The remaining two series
are by definition an Appel \hf{} $F_4$ with two variables
\cite{bateman}. This function, in the special case of $i=j=l=m=-1$
further reduces to Gauss' \hf{} ${}_2F_1$ \cite{bateman}.

In a manner similar to the previous results, there are solutions which have
three remaining variables, meaning that one of the series with unit argument is
summed out. There are six sets of these, determined by their variables, each
appearing four times, i.e., a fourfold degeneracy. Just to keep our accounting
straight, $4 \times 6 = 24$ systems yielding solutions with three variables.
These, added to the $36$ of the previous two-variables results, give us $60$
from the total of $81$ non-trivial systems.

The solutions within this category have functional dependencies given by
$(x,x,y),\;(z,z,y^{-1}), (x,y,y),\;(z,y^{-1},y^{-1}),$ 
$(x^{-1},x^{-1},z^{-1}),\;(z^{-1},z^{-1},x^{-1})$.

Note that the above triplets are conveniently arranged into pairs connected by
the symmetry $k^2 \leftrightarrow t^2$. 

Here we focus on the triple power series representation of interest, given by

\beq 
\label{S2} S_2^{AC} &=& P_2^{AC}
\sum_{n_1,n_2,n_3=0}^\infty  
\frac{(z)^{n_1+n_2}(y^{-1})^{n_3}}{n_1!n_2!n_3!}\frac{(m+\half
D|n_1)}{(1-i-j-\half D|n_1+n_2)}\nonumber \\[.25cm]
& \times &\frac{(l+\half D|n_2)(-i|n_1+n_2+n_3)(-\s|n_1+n_2+n_3)}{
(l+m+D|n_1+n_2)(1-i-l-m-D|n_3)} ,
\eeq
where 
\beq 
P_2^{AC} &=& \pi^D(t^2)^{\s}(-j|\s)(-l|l+m+\half D)(-m|l+m+\half D)\nonumber\\[.25cm] 
& \times &(l+m+D|i+j-l-m-\half D)(\s+\half D|-2\s-\half D).\nonumber
\eeq

Lastly, we consider solutions of systems of linear algebraic equations leading
to four variables, i.e., fourfold summation series with four variables. There
are $21$ of these, which completes the total of $81$ non-trivial solutions for
the systems. Again the functional variables are given paired with their
corresponding symmetries, as follows: $(z^{-1},z^{-1},x^{-1},x^{-1})$,
$(x,y,z,y^{-1})$, $(z,z,y^{-1},y^{-1}),\;(x,x,y,y)$,
$(y,y,z,y^{-1}),\;(y^{-1},y^{-1},x,y)$, $(x,y,x^{-1},z^{-1}),\;
(z,y^{-1},z^{-1},x^{-1})$, $(z,z,x^{-1},z^{-1}),\;
(x,x,z^{-1},x^{-1})$, $(z,y^{-1},z^{-1},z^{-1}),\;
(x,y,x^{-1},x^{-1})$.

To get the accounting straight, let us again mention that the first three appear
just once while the remaining nine appear twice, totalling the needed $21$ of
this category. Among them, there is a solution we want to pinpoint here, namely, 

\beq 
\label{S3} S_3^{AC} &=& P_3^{AC}
\sum_{\{n_i\}=0}^\infty  
\frac
{(z)^{n_1+n_2}(y^{-1})^{n_7+n_8}}{n_1!n_2!n_7!n_8!}\frac {(-j|n_1+n_2+n_7+n_8)(l+\half D|n_2+n_8)}
{(1-j+\s|n_7+n_8)} \nonumber \\[.25cm]
& \times & \frac{(m+\half D|n_1+n_7)}
{(1-i-j-\half D|n_1+n_2)} ,
\eeq 
where 
\beq 
P_3^{AC} &=& (k^2)^{\s}\;y^j\;(-i|-l-m-D)(-l|l+m+\half D)(-m|l+m+\half D) \nonumber\\
& \times & (\s+\half D|i+j-\s) .
\eeq 

\subsection{On-Shell Limit.}
 
Of course, particular cases of on-shell external legs must be contained in the
set of off-shell solutions. To check on this, let us take two legs on-shell,
namely, let $k^2=t^2=0$. Not all off-shell solutions $S^{AC}$ are suitable for
taking this particular limit, because some of them either vanish or are not defined in this regime. It is easy to see that such a suitable solution is given by
Eq.(\ref{S1}), because in this limit only the first term in the $F_4$ series is
non-zero while all the others vanish, leaving us with

\beq 
S_1^{AC}(k^2=t^2=0) &=& \pi^D(p^2)^\s (-i|\s)(-j|\s)(-l|-m-\half
D)(-m|l+m+\half D)\nonumber\\[.25cm]
& \times &(l+m+D|-l-\half D)(\s+\half D|-2\s-\half D).
\eeq 

This result is valid for arbitrary $D$ and {\em negative} exponents od
propagators. In order to confront this result with the known one we still need
to go further in specializing to the case where $i=j=l=m=-1$. Then, what we get
is the very result obtained by Hathrell \cite{hath} using standard procedures
for calculating Feynman diagrams in positive $D$. Of course, a more
straightforward way of getting this result using \ndim{} is to put the
corresponding legs on-shell from the very beginning, in Eq.(\ref{gauss}), and
what we get then is twelve systems to solve with non-trivial solutions, exactly
the number we have for the solution in question: a twelvefold degeneracy giving
the same result \cite{lab}.

We can also consider another special case, namely, the one where $p^2=k^2=0$.
This one is interesting because it contributes to another two-loop three-point
diagram \cite{kramer} if we apply the integration by parts technique
\cite{russo}. The general result, for arbitrary (negative) exponents of
propagators and positive dimension can be read off from the solution $S_2$,
Eq.(\ref{S2}),

\beq 
S_2^{AC}(p^2=k^2=0) &=& \pi^D(t^2)^{\s}(-j|\s)(-l|l+m+\half
D)(-m|l+m+\half D)\\[.25cm]  
& \times &(l+m+D|i+j-l-m-\half D)(\s+\half D|-2\s-\half D)\nonumber.
\eeq

Taking the same particular case of Kramer\cite{kramer} {\it et al}, i.e.,
$l=m=-2$ and $i=j=-1$, we obtain the well-known result in Euclidean space.

Two other simpler special cases can be read off from this graph, namely, when
$p=0,\; k^\nu=t^\nu$ (see Fig.2) and $k=0,\;p^\nu=-t^\nu$, as follows:

The solution $S_2$ gives us the first one,

\be 
S_2^{AC} (p=0,k^\nu=t^\nu) = (-i-j|i)(i+l+m+D|-i)P_2^{AC} .
\ee

Note that in the $n_1$ and $n_2$ series only the first term contributes and the
$n_3$ series reduces to a Gauss \hf{} with unit argument, thus a summable one.

The second case, $k=0,\;p^\nu=-t^\nu$, can be read off from $S_1$,

\be 
S_1^{AC}(k=0,p^\nu=-t^\nu) = \pi^D (p^2)^\s P_1^{AC} (-i+\s|-\s)(\half
D+j-\s|\s).
\ee

This result can be used to study two self-energy two-loop graphs and agrees
with our previously published results \cite{flying}.

Finally, let us check up on a solution that has four series with four
variables. Let $j=0$, so that we get the graph of Fig.3. The solution that
allows us to consider this limit is $S_3$,

\be 
S_3^{AC}(j=0) = \pi^D (k^2)^\s P_3^{AC}(j=0) .
\ee

Observe that there is no sum in the result. This is due to the factor
$(-j|n_1+n_2+n_7+n_8)$ which leads to only one non-vanishing term, i.e., when
$n_1=n_2=n_7=n_8=0$ in the series.

\section{Light-cone gauge loop integrals.}

For the vector gauge fields in the light-cone gauge \cite{leib} ghosts decouple
from the physical fields and for this reason the number of vertices we have in
the theory may be considerably reduced --- and consequently the number of
Feynman diagrams to deal with --- but the price we have to pay is that we are
left with a more complicated gauge boson propagator. This complexity manifests
itself in the form of a gauge dependent pole of the form $(k\cdot n)^{-1}$
where $n^{\mu}$ is a light-like vector which defines the gauge. It is a
well-known fact admit the light-cone experts that the use of Cauchy principal
value (CPV or PV for short) prescription to treat such a pole is plagued with
pathologies such as the impossibility of Wick rotation, the emergence of double
pole singularities at the one-loop level, and the incorrect exponentiation of
the Wilson loop\cite{wilson}, to name a few. To circumvent these difficulties, by the middle
of '80s, Mandelstam\cite{mandelstam} and Leibbrandt\cite{leib2} independently suggested prescriptions to
treat the so-called ``spurious'' singularities generated by the light-cone
propagator. Soon after it has been shown that both prescriptions were in fact
equivalent and became known as the ML prescription. Still later on the
prescription has been generalized to deal with generic non-covariant axial
gauges and sometimes it has been referred as the generalized Mandelstam-Leibbrandt prescription\cite{viena} in this
context. In parallel, it became clear that all those pathologies do arise
because the PV prescription violates causality and once causality is carefully
taken into account, no prescription is in fact needed \cite{pimentel}.

After these few words of elucidation to introduce the light-cone gauge to those
unfamiliar with it, we are in position to propose applying the \ndim{}
technology to see what happens in this case. We can anticipate, of course, 
some subtle intrinsic properties.

Just to make things easier, we follow the usual notation for the light-cone
that can be found in the majority of the specific literature on it.
 
First of all, let us consider the simplest of the one-loop integrals, 

\be 
L_1 = \int \frac{d^{2\omega}k}{k^2(k-b)^2(k\cdot n)} ,
\ee
which can be seen as the limiting case, i.e., $a \rightarrow 0$, of the more
general one, 

\be 
L_2 = \int \frac{d^{2\omega}k}{(k-a)^2(k-b)^2(k\cdot n)} .
\ee 

Although one can easily compute this last integral --- which arises in the
computation of one-loop four-point functions --- with the help of dimensional
regularization technique, the result of this particular integral is not
tabulated in the literature as far as we know it. Therefore it is a suitable
object to do our lab testing in \ndim{}, since the previous $L_1$ integral is
by far the most well-known integral in light-cone gauge.

Our starting point is again the Gaussian integral,

\beq 
\label{A}
{\cal A} &=& \int d^{2\omega}k\;
\exp{\left[-\alpha(k-a)^2-\beta(k-b)^2-\gamma(k\cdot n)\right]}
\nonumber\\
& = &\left(\frac{\pi}{\lambda}\right)^\omega \exp{\left\{-\frac{1}{\lambda}\left[ \alpha\beta (a-b)^2
-\alpha\gamma (a\cdot n)-\beta\gamma (b\cdot n) \right]\right\}}\nonumber\\
& = &\sum_{i,j,l=0}^\infty \frac{(-1)^{i+j+l}\alpha^i\beta^j\gamma^l}{i!j!l!} {\cal N}(i,j,l), \eeq
where $\lambda=\alpha+\beta$ and 
\be 
{\cal N}(i,j,l) = \int d^{2\omega}k \;(k-a)^{2i} (k-b)^{2j} (k\cdot n)^l .
\ee

Note that in the middle line of Eq.(\ref{A}) we are left with no factor
proportional to $n^2$ in the argument of the exponential , since it is zero in
the light-cone gauge  (by choice the $n^{\mu}$ vector is a light-like one.)
Also note that in \ndim{} scheme one has generic values of $(i,j,l)$ and not just
the specific quantum field theory values $i=j=l=-1$. 

By comparing the two expressions arising from the expansions of ${\cal A}$ we
obtain an expression in terms of multiple power series for the integral in
negative dimension, i.e.,

\beq 
{\cal N}(i,j,l) &=&
{\pi}^{\omega}(-1)^{i+j+l}{\Gamma}(1+i){\Gamma}(1+j){\Gamma}(1+l)\nonumber\\
&\times & \:{\sum^{\infty}_{n_i=0}}(-1)^{n_1+n_2+n_3}
         {\frac{(-n_1-n_2-n_3-{\omega})!}{n_1!n_2!n_3!n_4!n_5!}}
	 (a-b)^{n_1}(a\cdot n)^{n_2}(b\cdot n)^{n_3}\nonumber\\
&\times &\: {\delta}_{n_1+n_2+n_4,i}\:{\delta}_{n_1+n_3+n_5,j}\:
	 {\delta}_{n_2+n_3,l}\:{\delta}_{n_4+n_5,-(n_1+n_2+n_3+{\omega})}.
\eeq

The system here is far simpler than the former one in the two-loop covariant
case. There are altogether $5$ possible solutions but one of them is trivial.
Thus using the procedure outlined in the previous section and also in
\cite{lab} we construct two power series representations for the Feynman
integral in question. The solutions in which the sum indices $n_2$ and $n_5$
are left undetermined give us, after a suitable analytic continuation to
allow for {\em negative} values of $(i,j,l)$ and {\em positive} dimension,
\beq \label{nijl}
{\cal N}(i,j,l) &=& {\pi}^{\omega}[(a-b)^2]^{\rho-l}\left\{(b\cdot
n)^l Q_1^{AC}\; _2F_1(-l,{\omega}+j;1+j- \rho|z)\right. \nonumber\\  
&& + \left.(a\cdot n)^{\rho-j} (b\cdot n)^{-i-\omega} Q_2^{AC} \;{}_2F_1({\omega}+i,{\rho}+{\omega};1+{\rho}-j|z)
\right\},\eeq
where
\be
Q_1^{AC} = (-j|{\rho})(-i|i+j+{\omega})({\rho}+{\omega}|-2{\rho}-{\omega}+l),
\ee	
and
\be
Q_2^{AC} = (-j|i+j+{\omega})(-i|i+l-{\rho})(-l|j+l-{\rho}),	 
\ee
with $$z=\frac {a\cdot n}{b\cdot n},$$ and $\rho = i+j+l+\omega$. Letting $a=0$
Eq.(\ref{nijl}) gives the well-known result for $L_1$, see e.g. \cite{capper}.
If we want to write the result for $L_2$ in a more compact form we must rewrite
the \hf{}s $\F$ using a transformation formula \cite{bateman} of the type
$\F(...|x)\rightarrow\; \F(...|(1-x)^{-1})$. Considering
the special case where $i=j=l=-1$, we identify the sum as a single hypergeometric function, in Euclidean space,

\be
{\cal N}(-1,-1,-1)={\pi}^{\omega}
	{\frac{[(a-b)^2]^{{\omega}-2}}{(a.n)}}
	{\Gamma}(2-{\omega})B({\omega}-1,{\omega}-1)
	\ _2F_1(1,{\omega}-1;2{\omega}-2|u),
\ee
where $$u=\frac{(a-b)\cdot n}{b\cdot n}.$$

The remaining two other solutions for the system also lead to two Gaussian
\hf{}s but with $z^{-1}$ as variable. It is obtained from the former result by
making the simultaneous exchanges $a \leftrightarrow b$ and $i\leftrightarrow
j$ (which is an inherent symmetry of the referred integral.)

However, the above results for $L_1$ and $L_2$ are conspicuously pathological
in the sense that they violate causality, as discussed earlier on. In other
words, they are concordant with those results obtained in the PV prescription
scheme in positive dimension. This means that for each integral thus
calculated one has to subtract out the zero-mode contribution from it \cite{pimentel}. Yet we
would like to have causality preserving results without much ado. Is that
possible in the \ndim{} scheme? The answer is yes --- at least in principle ---
with some inherent subtleties as we shall shortly see.

Consider again the simplest of the one-loop light-cone integrals in the case we
have a two-degree violation of covariance, i.e., the tadpole-like integral with
tensorial structure,

\be
\label{tad}
T = \int \frac {d^{2\omega}k}{(k-p)^2}\frac {(k\cdot n^{\ast})}{(k\cdot
n)}.
\ee

Before we go on, a word of caution must be given just here. Note that the
structure of the integrand is important. In the light-cone gauge
Feynman-integrals, those factors bearing the external dual light-like vector
$n^{\ast}_{\mu}$ do {\em always} appear in the {\em numerator} of the
integrands. 

Therefore, our starting point to solve it in the \ndim{} scheme is,

\beq
{\cal B} &=&\int d^{2\omega}k\, \exp{\left[-\alpha (k-p)^2-\beta (k\cdot
n)-\gamma (k\cdot n^{\ast})\right]}\nonumber\\
&=&{\rm e}^{-\beta (p\cdot n)-\gamma (p\cdot n^{\ast})}\int d^{2\omega
}k\,{\rm e}^{-\alpha k^2-\beta (k\cdot n)-\gamma (k\cdot n^{\ast})}\nonumber\\
&=& \left ( \frac {\pi}{\alpha}\right)^{\omega} \exp \left\{-\beta (p\cdot n)-\gamma
(p\cdot n^{\ast}) + \frac {\beta\gamma (n\cdot
n^{\ast})}{2\alpha}\right\}\nonumber\\
&=&
\sum_{i,j,l=0}^{\infty}(-1)^{i+j+l}\frac{\alpha^i\beta^j\gamma^l}{i!j!l!}{\cal
T}.
\eeq
where ${\cal T}$ is the negative dimensional integral

\be
{\cal T} = \int d^{2\omega}k\:[(k-p)^2]^i\:(k\cdot n)^j\:(k\cdot n^{\ast})^l.
\ee

From the above equality, it is easy to get the result

\be
{\cal T} = (-\pi)^{\omega}\left (\frac {-2\:p\cdot n\; p\cdot
n^{\ast}}{n\cdot n^{\ast}}\right )^{i+\omega}(p\cdot n)^j (p\cdot
n^{\ast})^l\;\frac{(1-i-\omega|2i+\omega)}{(1+j|i+\omega)(1+l|i+\omega)}.
\ee

Now comes the crucial point. As mentioned earlier, the peculiarity of the
light-cone gauge is that the exponent $l \geq 0$ {\em always}. That means that
the Pochhammer's symbol containing it, namely $(1+l|i+\omega)$ must never be
analytic continued to allow for {\em negative} values of $l$. Bearing in mind
this restriction and subtlety, we proceed in the same manner as usual, to get

\be {\cal T}^{AC} = \pi^{\omega}\left (\frac {2\:p\cdot n\; p\cdot
n^{\ast}}{n\cdot n^{\ast}}\right )^{i+\omega}(p\cdot n)^j (p\cdot
n^{\ast})^l\;\frac{(-j|-i-\omega)}{(i+\omega|-2i-\omega)(1+l|i+\omega)}.
\ee

This is exactly the result we get through causal considerations (or by using
the ML prescription.) It is really quite an amazing result, since {\em no
prescription} has been called upon to deal with the so-called ``unphysical''
singularities characteristic of this gauge. The only consideration was that we
used the two-degree violation of covariance where both external vectors $n_\mu$
and its dual $n^{\ast}_\mu$ were treated in the same footing. It seems that
the outstanding property of translational invariance displayed by negative
dimensional integrals takes care of the causality principle naturally, a
principle that is required by {\em ad hoc} devised prescriptions
\cite{suzuki2}.

\section{Conclusion.}

We have shown in this paper how we can work out a two-loop vertex diagram with
all external legs off-shell using the \ndim{} technique to solve a pertinent
scalar Feynman integral. Altogether, twenty-one distinct results are obtained
for the considered integral. These are expressed in terms of power series that
can be identified as \hf{}s. The simples ones are Appel's $F_4$ \hf{} with two
variables, which for the particular cases where all the exponents of
propagators are set to minus one, can be transformed into even simpler ones of
the Gaussian $_2F_1$ \hf{} type. The technique also gives us several \ac{}
formulas between different results, because they arise from the same Feynman
integral (\ref{Indim}). 

The more outstanding results of this work, however, stems from our using
\ndim{} technology to evaluate the simplest one-loop light-cone integrals,
where  two distinct conclusions can be drawn: $(i)$ That when we consider
light-cone gauge with one-degree violation of covariance, i.e., consider only
the gauge-breaking external vector $n_\mu$, the results we get are concordant
with the usual PV prescription results, while $(ii)$ when we consider the
light-cone gauge with two-degree violation of covariance, where both vectors,
$n_\mu$ and its dual $n^{\ast}_\mu$ are treated on the same footing, the
result we get is concordant with that obtained via causal considerations (or
equivalently the use of the ML prescription.) A more outstanding conclusion we
draw from this last result is that \ndim{}, somehow, in a manner that we still
do not understand clearly, takes care naturally of the causality principle that
should constrain all transition amplitudes. Moreover, that which in times past
required an {\em ad hoc} prescription to deal with gauge dependent poles is
with much finesse avoided by \ndim.

\acknowledgments{
A.G.M.S. and R.B. gratefully acknowledge FAPESP (Funda\c c\~ao de Amparo \`a Pesquisa do
Estado de S\~ao Paulo, Brasil) for financial support. }

\end{document}